\documentstyle[preprint,aps]{revtex}
\tightenlines
\begin{document}
\draft
\title
{Hausdorff dimension, anyonic distribution functions, and duality}
\author
{Wellington da Cruz\footnote{E-mail: wdacruz@exatas.campus.uel.br}} 
\address
{Departamento de F\'{\i}sica,\\
 Universidade Estadual de Londrina, Caixa Postal 6001,\\
Cep 86051-990 Londrina, PR, Brazil\\}
\date{\today}
\maketitle
\begin{abstract}
We obtain the distribution functions for anyonic 
excitations classified 
into equivalence 
classes labeled by Hausdorff dimension $h$ and 
as an example 
of such anyonic systems, we consider the 
collective excitations of the Fractional Quantum 
Hall Effect ( FQHE ).
We also introduce the concept of duality 
between such classes, 
defined by $\tilde{h}=3-h$. In this way, we confirm that 
the filling factors for which 
the FQHE were 
observed just appears into these classes and 
the internal duality for a given class $h$ or $\tilde{h}$ 
is between quasihole and 
quasiparticle excitations for these FQHE systems. 
Exchanges of dual 
pairs $\left(\nu,\tilde{\nu}\right)$, suggests 
conformal invariance. A connection between equivalence classes $h$ 
and the modular group for the quantum phase transitions of the FQHE 
is also obtained. A $\beta-$function is also defined for the complex 
conductivity which embodies the $h$ classes.

\end{abstract}



We have obtained in\cite{R1} from a continuous family of 
Lagrangians for fractional spin particles a path integral 
representation for the propagator and its representation in 
moment space. On the other hand, the trajectories swept out 
by scalar and spinning particles can be characterized by 
the fractal parameter $h$ ( or the Hausdorff dimension ). 
We have 
that $L$, 
the length of closed trajectory with size $R$ has its fractal 
properties described by $L\sim R^h$\cite{R2}, such that 
for scalar particle 
 $L\sim \frac{1}{p^2}$, $R^2\sim L$ and $h=2$ ; for  
 spinning particle, $
 L\sim \frac{1}{p}$, $R^1\sim L$ and $h=1$. From the form of 
 anyonic propagator given in\cite{R1}, with the spin defined 
 into the interval $0\leq s\leq 0.5$, we have extracted the 
 following formula $h=2-2s$, which relates the Hausdorff 
 dimension $h$ and the spin $s$ of the particle. Thus, 
 for anyonic particle, $h$ takes values within the interval 
 $1$$\;$$ < $$\;$$h$$\;$$ <$$\;$$ 2$. 
 In \cite{R3}, we have classified 
 the fractional 
spin particles or anyonic excitations in terms of 
equivalence classes labeled by $h$. 
Therefore, such particles in a specific class 
can be considered on equal footing.

On the other hand, in the context of FQHE systems, the 
filling factor, 
a parameter which characterize that phenomenon, 
can be classified in this 
terms too. We have, therefore, a new
hierarchy scheme for the filling factors\cite{R4}, which is 
extracted from the relation between $h$ and the 
statistics $\nu$ 
( or the filling factors ). Our approach, contrary to 
literature\cite{R5} do not have an empirical character 
and we 
can predicting for which values of $\nu$ FQHE can be 
observed. 
A braid group structure behind this classification 
also was 
noted\cite{R6} and in\cite{R3} a topological invariant 
${\cal W}=h+2s-2p$, was introduced which relates 
a characteristic 
of the multiply connected spaces with the numbers of 
fractional spin particles $p$, and the quantities 
related 
to the particles, $h$ and $s$ ( note that this invariant
 has a connection with the {\it Euler characteristic} of the 
 surface, and hence we have a physical meaning 
 for this number ). The anyonic model, a 
charge-flux 
system, constitutes a topological obstruction 
( holes with hard core repulsion ) in this description and 
the elements of the braid group are 
equivalents trajectories. 
 
 Now, we propose a statistical weight for 
such excitations in terms 
of $h$, as follows:
  
\begin{equation}
\label{e.1}
\omega_{j}=\frac{\left[G_{j}+(N_{j}-1)(h-1)\right]!}{N_{j}!
\left[G_{j}+(N_{j}-1)(h-1)-N_{j}\right]!},
\end{equation}

\noindent where $G_{j}$ means a group of quantum states, 
$N_{j}$ is the number 
of particles and for, $h=2$ we have bosons and for 
$h=1$ we have fermions. For 
$1$$\;$$ < $$\;$$h$$\;$$ <$$\;$$ 2$, we have anyonic 
excitations which interpolates between these two extremes. 
The fractal parameter $h_{i}$ 
 ( $i$ means a specific interval ) is related 
 to statistcs $\nu$, in the 
  following way:  

\begin{eqnarray}
\label{e.2}
&&h_{1}=2-\nu,\;\;\;\; 0 < \nu < 1;\;\;\;\;\;\;\;\;
 h_{2}=\nu,\;\;\;\;
\;\;\;\;\;\;\;\;\; 1 <\nu < 2;\;\nonumber\\
&&h_{3}=4-\nu,\;\;\;\; 2 < \nu < 3;\;\;\;\;\;\;\;\;
h_{4}=\nu-2,\;\;\;\;\;\;\; 3 < \nu < 4;\;\nonumber\\
&&h_{5}=6-\nu,\;\;\;\; 4 < \nu < 5;\;\;\;\;\;\;\;
h_{6}=\nu-4,\;\;\;\;\;\;\;\; 5 < \nu < 6;\;\\
&&h_{7}=8-\nu,\;\;\;\; 6 < \nu < 7;\;\;\;\;\;\;\;
h_{8}=\nu-6,\;\;\;\;\;\;\;\; 7 < \nu < 8;\;\nonumber\\
&&h_{9}=10-\nu,\;\;8 < \nu < 9;\;\;\;\;\;\;
h_{10}=\nu-8,\;\;\;\;\;\; 9 < \nu < 10;\nonumber\\
&&etc,\nonumber
\end{eqnarray}

\noindent such that this spectrum of $\nu$, as we can see, 
has a 
complete mirror symmetry and for a given $h$, we collect
 different values of $\nu$ in a specific class, for example, 
 consider $h=\frac{5}{3}$ and $h=\frac{4}{3}$, so we obtain;
 $\left\{\frac{1}{3},\frac{5}{3},\frac{7}{3},
 \frac{11}{3},\frac{13}{3},\frac{17}{3},\cdots\right\}_
{h=\frac{5}{3}}$ and  $\left\{\frac{2}{3},
\frac{4}{3},\frac{8}{3},\frac{10}{3},\frac{14}{3},
\frac{16}{3},
\cdots\right\}_
{h=\frac{4}{3}}$.

These particles or excitations, in the class, share some 
common 
characteristics, according our interpretation, in the 
same way that bosons and fermions. We observe that 
each interval 
contributes with one and only one particle to the 
class and the 
statistics and the spin, are related by $\nu=2s$. 

We stress that, our expression for statistical 
weight 
Eq.(\ref{e.1}), is more general than an expression given 
in\cite{R7},
once if we consider the relation between $h$ and $\nu$, we 
just obtain that first result and extend it for the 
complete spectrum of statistics $\nu$, for example: 
 
\begin{eqnarray}
\label{e.4}
&&\omega_{j}=\frac{\left[G_{j}+(N_{j}-1)(1-\nu)\right]!}
{N_{j}!\left[G_{j}+(N_{j}-1)(1-\nu)-N_{j}\right]!},
 \;\;\;\; 0 < \nu < 1; \nonumber\\
&&\omega_{j}=\frac{\left[G_{j}+(N_{j}-1)(\nu-1)\right]!}
{N_{j}!\left[G_{j}+(N_{j}-1)(\nu-1)-N_{j}\right]!},
\;\;\;\; 1 < \nu < 2; \\
&&\omega_{j}=\frac{\left[G_{j}+(N_{j}-1)(3-\nu)\right]!}
{N_{j}!\left[G_{j}+(N_{j}-1)(3-\nu)-N_{j}\right]!},
\;\;\;\; 2 < \nu < 3;\nonumber\\
&&\omega_{j}=\frac{\left[G_{j}+(N_{j}-1)(\nu-3)\right]!}
{N_{j}!\left[G_{j}+(N_{j}-1)(\nu-3)-N_{j}\right]!},
\;\;\;\; 3 < \nu < 4;\nonumber\\
&&etc.\nonumber
\end{eqnarray}

\noindent  The expressions 
Eq.(\ref{e.4}) were possible because of the mirror symmetry 
as just have said above. But, our approach 
in terms of Hausdorff dimension $h$, have an advantage, 
because we collect into equivalence class the anyonic 
excitations and so, we consider on equal footing the 
excitations 
in the class. We have, therefore, a new approach 
for fractional spin particles. For different 
species of particles, the statistical weight take the form

\begin{equation}
\label{e.5}
\Gamma=\prod_{j}\omega_{j}=\prod_{j}\frac{\left
[G_{j}+(N_{j}-1)
(h-1)\right]!}{N_{j}!\left[G_{j}+(N_{j}-1)(h-1)-N_{j}\right]!},
\end{equation}
\noindent which reduces to Eq.(\ref{e.1}) for 
only one species. Now, we can consider the entropy 
for a given class $h$. Taking the logarithm of 
Eq.(\ref{e.5}), 
with the condition that, $N_{j}$ and $G_{j}$ are vary 
large numbers and 
defining the average occupation numbers, $n_{j}=
\frac{N_{j}}{G_{j}}$, 
we obtain for a gas not in equilibrium, an expression 
for the entropy

\begin{eqnarray}
&&{\cal S}=\sum_{j}G_{j}\left\{\left[1+n_{j}(h-1)
\right]\ln\left(
\frac{1+n_{j}(h-1)}{1+n_{j}(h-1)-n_{j}}\right)\right.\\
&&-\left.n_{j}\ln\left(\frac{n_{j}}{1+n_{j}(h-1)-n_{j}}
\right)\right\},
\nonumber
\end{eqnarray}

\noindent such that for $h=2$ and $h=1$, we obtain
the expressions for a Bose and Fermi 
gases not in equilibrium\cite{R8}, 
  
\begin{eqnarray}
&&{\cal S}=\sum_{j}G_{j}\left\{\left(1+n_{j}\right)\ln
\left(1+n_{j}\right)-n_{j}\ln n_{j}\right\};\\
&&{\cal S}=-\sum_{j}G_{j}\left\{n_{j}\ln n_{j}+
\left(1-n_{j}\right)\ln \left(1-n_{j}\right)\right\},
\end{eqnarray}
\noindent respectively.

The distribution function for a gas of the class $h$ can 
be obtained from the condition of the entropy be a 
maximum. Thus, 
we have

\begin{eqnarray}
\label{e.9}
n_{j}\xi=\left\{1+n_{j}(h-1)\right\}^{h-1}\left\{1+n_{j}
(h-2)\right\}^{2-h}
\end{eqnarray}

\noindent or 
\begin{eqnarray}
\label{e.45} 
n_{j}=\frac{1}{{\cal{Y}}(\xi)-h},
\end{eqnarray}

\noindent where the function ${\cal{Y}}(\xi)$ satisfies 
\begin{eqnarray}
\xi=\left\{{\cal{Y}}(\xi)-1\right\}^{h-1}
\left\{{\cal{Y}}(\xi)-2\right\}^{2-h},
\end{eqnarray}
 
 \noindent and $\xi=\exp\left\{(\epsilon_{j}-\mu)/KT\right\}$, 
has the usual definition. The Bose and Fermi distributions
 are obtained for values of $h=2,1$ respectively. 
 At this point, 
 we observe that the condition of periodicity on 
 the statistics $\nu$, in our 
 approach, is expressed as 
 
 \begin{equation}
 n_{j}(\nu)=n_{j}(\nu+2),
 \end{equation}
 
 \noindent and the equivalence classes $h$ always respect 
 this condition naturally. 
 On the other hand, this also means that, as $\nu=2s$, particles 
 with distinct values of spin $s$ into the class $h$ obye a 
 specific fractal statistics Eq.(\ref{e.45}).

In the 
context of the 
Fractional Quantum Hall Effect ( FQHE ), as we said 
in the introduction, 
the filling factor ( rational number with an 
odd denominator ), 
can be also classified in 
terms of $h$\cite{R4}. We have that the anyonic excitations 
are collective 
excitations manifested as quasiparticles or quasiholes in FQHE 
systems. 
Thus, for example, we have the collective 
excitations as given in 
the beginning for $h=\frac{5}{3}$ and 
$h=\frac{4}{3}$. Now, we 
have noted 
that these collections contain filling fractions, in 
particular, $\nu=
\frac{1}{3}$ and $\nu=\frac{2}{3}$, that experimentally 
were observed\cite{R9} and so we are 
able to estimate for which values of $\nu$ the largest 
charge gaps occurs, or alternatively, we can {\it predicting} 
FQHE. As was 
observed in\cite{R4} this is a {\bf new hierarchy scheme} for 
the filling 
factors, that {\bf expresses the occurrence of the} FQHE {\bf in more 
fundamental terms}, that is, {\bf relating the fractal parameter} $h$ 
{\bf and the filling factors} $\nu$. Of course, in our approach 
we do not have empirical expressions like this one\cite{R10}, 
$\nu=\frac{n}{2pn\pm 1}$, for the experimental occurrence 
of FQHE. After all, for 
anyonic excitations in a stronger 
magnetic field 
at low temperature, our results 
( Eq.\ref{e.4} ) obtained via an approach completely 
distinct of\cite{R7} can be 
considered. 

In another way, we introduce the concept of duality between 
equivalence classes, defined by 

\begin{equation} 
\label{e.10}
\tilde{h}=3-h,
\end{equation}

\noindent such that, for $h=1$, $\tilde{h}=2$ and 
for $h=2$, $\tilde{h}=1$, that is fermions and bosons 
are dual objects. This means that they can be 
considered supersymmetric particles\cite{R11}. 

For a set of values of the filling factors 
$\nu$ experimentally observed\cite{R9}, second 
our relations ( Eq.\ref{e.2} and Eq.\ref{e.10} ), 
we get the 
classes, $h$ and $\tilde{h}$:

\begin{eqnarray}
&&\left\{\frac{1}{3},\frac{5}{3},\frac{7}{3},
 \frac{11}{3},\cdots\right\}_
{h=\frac{5}{3}},\;\;\;\;\;\;\;\;\;\;\left\{\frac{2}{3},
\frac{4}{3},\frac{8}{3},\frac{10}{3},\cdots\right\}_
{{\tilde{h}}=\frac{4}{3}};\nonumber\\
&&\left\{\frac{1}{5},\frac{9}{5},\frac{11}{5},
 \frac{19}{5},\cdots\right\}_
{h=\frac{9}{5}},\;\;\;\;\;\;\;\;\left\{\frac{4}{5},
\frac{6}{5},\frac{14}{5},\frac{16}{5},\cdots\right\}_
{{\tilde{h}}=\frac{6}{5}};\nonumber\\
&&\left\{\frac{2}{7},\frac{12}{7},\frac{16}{7},
 \frac{26}{7},\cdots\right\}_
{h=\frac{12}{7}},\;\;\;\;\;\left\{\frac{5}{7},
\frac{9}{7},\frac{19}{7},\frac{23}{7},\cdots\right\}_
{{\tilde{h}}=\frac{9}{7}};\nonumber\\
&&\left\{\frac{2}{9},\frac{16}{9},\frac{20}{9},
 \frac{34}{9},\cdots\right\}_
{h=\frac{16}{9}},\;\;\;\;\;\left\{\frac{7}{9},
\frac{11}{9},\frac{25}{9},\frac{29}{9},\cdots\right\}_
{{\tilde{h}}=\frac{11}{9}};\nonumber\\
&&\left\{\frac{2}{5},\frac{8}{5},\frac{12}{5},
 \frac{18}{5},\cdots\right\}_
{h=\frac{8}{5}},\;\;\;\;\;\;\;\;\left\{\frac{3}{5},
\frac{7}{5},\frac{13}{5},\frac{17}{5},\cdots\right\}_
{{\tilde{h}}=\frac{7}{5}};\\
&&\left\{\frac{3}{7},\frac{11}{7},\frac{17}{7},
 \frac{25}{7},\cdots\right\}_
{h=\frac{11}{7}},\;\;\;\;\;\left\{\frac{4}{7},
\frac{10}{7},\frac{18}{7},\frac{24}{7},\cdots\right\}_
{{\tilde{h}}=\frac{10}{7}};\nonumber\\
&&\left\{\frac{4}{9},\frac{14}{9},\frac{22}{9},
 \frac{32}{9},\cdots\right\}_
{h=\frac{14}{9}},\;\;\;\;\;\left\{\frac{5}{9},
\frac{13}{9},\frac{23}{9},\frac{31}{9},\cdots\right\}_
{{\tilde{h}}=\frac{13}{9}};\nonumber\\
&&\left\{\frac{6}{13},\frac{20}{13},\frac{32}{13},
 \frac{46}{13},\cdots\right\}_
{h=\frac{20}{13}},\;\;\;\left\{\frac{7}{13},
\frac{19}{13},\frac{33}{13},\frac{45}{13},\cdots\right\}_
{{\tilde{h}}=\frac{19}{13}};\nonumber\\
&&\left\{\frac{5}{11},\frac{17}{11},\frac{27}{11},
 \frac{39}{11},\cdots\right\}_
{h=\frac{17}{11}},\;\;\;\left\{\frac{6}{11},
\frac{16}{11},\frac{28}{11},\frac{38}{11},\cdots\right\}_
{{\tilde{h}}=\frac{16}{11}};\nonumber\\
&&\left\{\frac{7}{15},\frac{23}{15},\frac{37}{15},
 \frac{53}{15},\cdots\right\}_
{h=\frac{23}{15}},\;\;\;\left\{\frac{8}{15},
\frac{22}{15},\frac{38}{15},\frac{52}{15},\cdots\right\}_
{{\tilde{h}}=\frac{22}{15}}.\nonumber
\end{eqnarray}

\noindent We {\it emphasize} that in each class, some 
filling factors are just the experimental values 
observed, 
that is, the Hall resistance  develops plateaus in these 
quantized values, which are related to the fraction of 
electrons that form collective excitations as 
quasiholes or quasiparticles in FQHE systems. The 
relation 
of duality between equivalence classes labeled by $h$ 
can, therefore, indicates a way as determine the dual 
of a specific value of $\nu$ ( or ${\tilde{\nu}}$ ) 
observed. Note also that the class $
\left\{1,3,5,7,\cdots\right\}_{h=1}$ 
gives us the odd filling factors 
for the integer quantum Hall effect. An interpretation 
of the fractal parameter $h$ as some kind of {\it 
order parameter} which characterizes the ocurrence of FQHE 
is enough attractive. We can say that the 
concept of {\it duality} connects a {\it 
quasi-bosonic regime} $h\sim 2$ to a {\it 
quasi-fermionic regime} $h\sim 1$\cite{R13}, as $(h,\nu)=
(\frac{5}{3},\frac{1}{3})$ to $(\frac{4}{3},\frac{2}{3})$. 
Another view is the relation $L\sim R^h$, which shows us a 
{\it scaling law} behind this characterization of the FQHE. 

We also observe that our approach, in terms of equivalence 
classes for the filling factors, embodies 
the structure of the {\it modular group} as discussed in\cite{R11}. For 
that, we consider the sequences given by Dolan\cite{R11} as

\begin{eqnarray}
&&(h,\nu)=\left(\frac{5}{3},\frac{1}{3}\right)\rightarrow \left(\frac{8}{5},
\frac{2}{5}\right)
\rightarrow \left(\frac{11}{7},\frac{3}{7}\right)\rightarrow 
\left(\frac{14}{9},\frac{4}{9}\right)
\rightarrow \left(\frac{17}{11},\frac{5}{11}\right)\rightarrow 
\left(\frac{20}{13},\frac{6}{13}\right) \rightarrow \cdots;\nonumber\\
&&(h,\nu)=\left(\frac{19}{13},\frac{7}{13}\right)\rightarrow 
\left(\frac{16}{11},\frac{6}{11}\right)
\rightarrow \left(\frac{13}{9},\frac{5}{9}\right)
\rightarrow \left(\frac{10}{7},\frac{4}{7}\right)
\rightarrow \left(\frac{7}{5},\frac{3}{5}\right) \rightarrow 
\left(\frac{4}{3},\frac{2}{3}\right)\rightarrow (1,1);\\
&&(h,\nu)=\left(\frac{4}{3},\frac{2}{3}\right)\rightarrow 
\left(\frac{9}{7},\frac{5}{7}\right)
\rightarrow \left(\frac{14}{11},\frac{8}{11}\right)\rightarrow 
\left(\frac{19}{15},\frac{11}{15}\right)\rightarrow\cdots;\nonumber\\
&&(h,\nu)=\left(\frac{7}{5},\frac{3}{5}\right)\rightarrow 
\left(\frac{4}{3},\frac{2}{3}\right)
\rightarrow \left(\frac{9}{7},\frac{5}{7}\right)\rightarrow\cdots;\nonumber
\end{eqnarray} 

\noindent and we can verify that the other 
sequences follow according 
$h$ increases or decreases within the interval 
$1$$\;$$ < $$\;$$h$$\;$$ <$$\;$$ 2$, respecting $\nu$ 
entries in each class. The transitions allowed are 
those generated by the condition $\mid p_{2}q_{1}-p_{1}q_{2}\mid=1$, 
with $h_{1}=\frac{p_{1}}{q_{1}}$ and $h_{2}=\frac{p_{2}}{q_{2}}$. 
Thus our formulation satisfies in the same 
way the constraint given for the filling factors $\nu$.

We have that the transition between two Hall plateaus 
is a quantum phase 
transition generated by quantum fluctuations when 
the external magnetic 
field is varied, thus we have a {\it correlation length}
 ( remember $L\sim R^h$ ) given by

\begin{equation}
\zeta\sim\frac{1}{{\mid l-l_{c}\mid}^{h}},
\end{equation}

\noindent where $l$ is the {\it magnetic length} which depends on 
magnetic field and temperature. We take $h=2$, because 
we consider the {\it Hall 
fluid as a bosonic fluid} in according to diverse
field-theoretical models 
in the literature for FQHE\cite{R11} and better, 
experimentally the exponent of $\Delta{l}=l-l_{c}$ 
is equal to $2.02$. Also 
the dimensionless parameter $u=\frac{n\;e\;\Delta{h}}{T^{\mu}}$ ( 
$\Delta{h}=h_{1}-h_{2}$,$\;$ $n$ is the density 
of particles with charge $e$ ) 
can be used to describe the crossover between two Hall plateaus
, with the 
{\it critical exponent} $\mu=\frac{1}{h}=0.5$ 
( assumed universal ), in according 
with the experimental value $0.45\pm 0.05$. In this way we 
define $\beta-$function which describes just the 
crossover between two Hall plateaus as a complex 
analytic function of the complex conductivity

\begin{equation}
\beta(\sigma)=\frac{d\;\sigma}{d\;s},
\end{equation}

\noindent where $\sigma$ is the complex 
conductivity and $s$ is a real analytic monotonic function of $u$.

Following Dolan\cite{R11}, we can define

\begin{equation}
\sigma(\Delta{h})=\frac{p_{2}q_{2}\left\{K^{\prime}(w)\right\}^2+
p_{1}q_{1}\left\{K(w)\right\}^2+\imath K^{\prime}(w)K(w)}
{q_{1}^2\left\{K(w)\right\}^2+q_{2}^2\left\{K^{\prime}(w)\right\}^2},
\end{equation}

\noindent where $K^{\prime}(w)$ and $K(w)$ are complete 
elliptic integrals of the second kind, with 

\begin{equation}
w^2=\frac{1}{2}\left\{1+sign(\Delta{h})\sqrt{1-e^{-\left(\frac{A\Delta{h}}
{\eta(\Delta{h})T^{\mu}}\right)^2}}\right\},
\end{equation}

\noindent with the complementary modulus $w^{\prime}$ defined as
 ${w^{\prime}}^2=1-w^2$, 
$K^{\prime}(w)=K(w^{\prime})$ and 
$\mu=\frac{1}{h}=0.5$ ( Hall fluid is bosonic ), 
$A$ is a positive real constant 
( which can depend on eletron and impurity density, or other parameters )
 and the linear function $
\eta(\Delta{h})=\alpha\left\{(q_{1}-q_{2})\Delta{h}+\alpha\right\}$, 
$\alpha=p_{2}-p_{1}-(q_{2}-q_{1})h_{c}$ and $h_{c}=h_{2}$ is the class 
of the filling factor $\nu_{c}=\nu_{2}$ at the critical 
value of the magnetic field.

Now, in another way, we give more arguments in favour of our approach. 
We have that for excitations above the 
Laughlin ground state, the exchange of two 
quasiholes\cite{R12} with coordinates $z_{\alpha}$ and 
$z_{\beta}$ produces the condition on the phase

\begin{equation}
\label{e.23}
\exp\left\{\imath\pi \nu_{1}\right\}=
\exp\left\{\imath\pi\frac{1}{m}\right\},
\end{equation}
\noindent with $\nu_{1}$$=$$\frac{1}{m}+2p_{1}$; 
and for a second generation 
of quasihole excitations, the effective wavefunction 
carries the factor $\left(z_{\alpha}-z_{\beta}
\right)^{\nu_{2}}$, 
with $\nu_{2}$$=$$\frac{1}{\nu_{1}}+2p_{2}$; 
$m=3,5,7,\cdots$ 
and $p_{1}$,$\;$$p_{2} $ are positive integers. 
In \cite{R6} we have noted that these conditions 
over the filling factor $\nu$ confirms our 
classification of the collective excitations in terms 
of $h$. Another 
interesting point is that in each class, we have more 
filling factors 
which those generates by ( Eq.\ref{e.23} ), that is, 
our classification 
cover a more complete spectrum of states. Now, 
we can see that the duality between equivalence classes 
means also duality between quasiholes and 
quasiparticles. There exist 
internal and external dualities. This means 
that for charges 
${\cal{Q}}=\pm \nu e $, we have an internal 
duality in each 
equivalence class $h$ or $\tilde{h}$. External 
duality stand 
for dual classes, 
 $h$ and $\tilde{h}$. Therefore, as we 
said elsewhere\cite{R13}, $h$ {\it tells us about the 
nature of the anyonic excitations}.

On the other hand, we note that the anyonic exchanges of 
dual get 
a phase difference, modulo constant,

\begin{equation}
\left|\nu-\tilde{\nu}\right|=\left|\Delta\nu\right|=\left|h-
\tilde{h}\right|=const,
\end{equation}

\noindent suggesting an invariance, conformal symmetry. 
We have also for 
the elements of $h$ and $\tilde{h}$, the following relations

\begin{eqnarray}
\label{e.19}
&&\frac{\nu_{i+1}-\nu_{i}}{\tilde{\nu}_{j+1}
-\tilde{\nu}_{j}}=1;\\
&&\frac{\nu_{j+1}-\nu_{j}}{\tilde{\nu}_{i+1}
-\tilde{\nu}_{i}}=1,\nonumber
\end{eqnarray}

\noindent with $i=1,3,5,etc.$ and $j=2,4,6,etc.$; 
such that the pairs 
$\left(i,j\right)=\left(1,2\right),\left(3,4\right),
etc.$ satisfy the 
expressions ( Eq.\ref{e.19} ). A discussion about 
the thermodynamics for 
fractal statistics was started in\cite{R14}.

In summary, we have obtained distribution 
functions ( Eq.\ref{e.9} ) for 
anyonic excitations in terms of the Hausdorff dimension 
$h$, which classifies the anyonic excitations into 
equivalence classes and reduces to 
fermionic and bosonic distributions, when $h=1$ and 
$h=2$, respectively. This constitutes a new approach 
for such systems of fractional spin particles. 
In this way, we have extended
 ( Eq.\ref{e.4} ) results 
of the literature\cite{R7} 
for the complete spectrum of 
statistics $\nu$. A connection with the FQHE, 
considering the filling factors into equivalence classes 
labeled by $h$ was also
considered and an estimate for occurrence 
of FQHE made. A connection 
between equivalence classes $h$ and the modular group for 
the quantum phase transitions of the FQHE was also noted. 
A $\beta-$function is write down which embodies $h$ classes. The 
concept of duality between equivalence classes confirms that 
all this work.


\begin{thebibliography}{99}
\bibitem{R1} W. da Cruz, preprint/UEL-DF/W-01/97, 
hep-th/9803020. 
\bibitem{R2} A. M. Polyakov, in {\it Proc. Les
 Houches Summer School
 {\bf vol. IL}}, ed. E. Br\'ezin and J. Zinn-Justin
  (North Holland, 1990), 305. 
\bibitem{R3} W. da Cruz, preprint/UEL-DF/W-02/97, 
hep-th/9803043; 
ibid, preprint/UEL-DF/W-03/97, hep-th/9802135.
\bibitem{R4} W. da Cruz, preprint/UEL-DF/W-04/97, 
cond-mat/9802266.
\bibitem{R5} F. D. M. Haldane, Phys. Rev. Lett. 
{\bf 51}, 605 (1983); B. I. Halperin, Phys. Rev. Lett. 
{\bf 52}, 1583 (1984); see also\cite{R9}.
\bibitem{R6} W. da Cruz, preprint/UEL-DF/W-05/97, 
cond-mat/9802267.   
\bibitem{R7} Y. S. Wu, Phys. Rev. Lett. {\bf 73}, 
922 (1994); 
S. B. Isakov, Mod. Phys. Lett. {\bf B8}, 319 (1994); 
A. K. Rajagopal, Phys. Rev. Lett. {\bf 74}, 1048 (1995).
\bibitem{R8} E. M. Lifshitz and L. P. Pitaevskii, 
{\it Statistical Physics, Part 1, 
3rd. edition } ( Pergamon Press, Oxford, 1980 ).
\bibitem{R9} A. H. MacDonald, cond-mat/9410047; 
C. Gros and A. H. MacDonald, 
Phys. Rev. {\bf B42}, 10811 (1990) and 
reference therein; A. M. Chang in {\it The 
Quantum Hall Effect} 
( Springer Verlag, 1990 ), Ed. by R. E. 
Prange and S. M. Girvin
and references therein; {\it Perspectives 
in Quantum Hall Effects}, Ed. by 
Sankar Das Sarma and Aron Pinczuk ( Wiley, New York, 1997 ); 
T. Chakraborty and P. Pieti\"ainen, 
{\it The Fractional 
Quantum Hall Effect: Perspectives of an 
incompressible quantum fluid} 
( Springer-Verlag, New York, 1988 ), 
Springer Series in Solid State Sciences, {\bf 85}; {\it 
Quantum Hall Effect: A Perspective }, Ed. by A. H. 
MacDonald ( Kluwer, Boston, 1989 ). 
\bibitem{R10} J. K. Jain and R. K. Kamilla, cond-mat/9704031.
\bibitem{R11} After submit this paper, I found 
in the literature different formulations relating 
duality and FQHE: 
S. Kivelson, D-H. Lee and S-C. Zhang, 
Phys. Rev. {\bf B46}, 2223 (1992);
A. P. Balachandran, L. Chandar and 
B. Sathiapalan, Nucl. Phys. {\bf B443}, 465 (1995); E. 
Shimshoni, S. L. Sondhi and D. Shahar, cond-mat/9610102; 
E. Fradkin and S. A. Kivelson, Nucl. Phys. 
{\bf B474}, 543 (1996); 
A. Cappelli and G. R. Zemba, Nucl. Phys. 
{\bf B490}, 595 (1997); 
T. Gannon, Nucl. Phys. {\bf B491}, 659 (1997); 
C. P. Burgess and C. A. L\"utken, Nucl. Phys. 
{\bf B500}, 367 (1997); 
P. Degiovanni, C. Chaubet and R. Melin, cond-mat/9711173; 
B. P. Dolan, cond-mat/9805171, cond-mat/9809294, hep-th/9811218; 
S. Skoulakis and S. Thomas, cond-mat/9806051,  
and references therein. For an experimental accounts about 
Hall transitions ( correlation length and 
critical exponent ) see: D. Shahar et al, Phys. Rev. 
Lett. {\bf 79}, 479 (1997); 
R. T. F. van Schaijk et al, cond-mat/9812035.
\bibitem{R12} A. Lerda, {\it Anyons}, Lectures 
Notes in Physics 
( Springer Verlag, 1992 ); R. B. Laughlin, Phys. Rev. Lett. 
{\bf 50}, 1395 ( 1983 ); ibid, Phys. Rev. {\bf B23}, 3383 
(1983).
\bibitem{R13} W. da Cruz, preprint/UEL-DF/W-03/97, 
hep-th/9802135; ibid, preprint/UEL-DF/W-04/97, 
cond-mat/9802266. 
\bibitem{R14} W. da Cruz, preprint/UEL-DF/981004, 
cond-mat/9811007.
\end{thebibliography}
\end{document}